\documentclass[aps,prb,preprint,groupedaddress]{revtex4-2}

\usepackage{amsmath}
\usepackage{graphicx}
\usepackage{subcaption}

\newcommand{\mean}[1]{\left\langle #1 \right\rangle}

\newcommand{\ket}[1]{| #1 \rangle}

\begin{document}

\preprint{}

\title{Melting of atomic hydrogen and deuterium with path-integral Monte Carlo}


\author{Kevin K. Ly}

\author{David M. Ceperley}
\affiliation{Department of Physics, University of Illinois at Urbana-Champaign}


\date{\today}

\begin{abstract}
    We calculate the melting line of atomic hydrogen and deuterium up to 900 GPa with path-integral Monte Carlo using a machine-learned interatomic potential.
    We improve upon previous simulations of melting by treating the electrons with reptation quantum Monte Carlo, and by performing solid and liquid simulations using isothermal-isobaric path-integral Monte Carlo.
    The resulting melting line for atomic hydrogen is higher than previous estimates.
    There is a small but resolvable decrease in the melting temperature as pressure is increased, which can be attributed to quantum effects.
\end{abstract}


\maketitle


\section{\label{s:intro}Introduction}

The synthesis of solid atomic hydrogen has been a decades-long pursuit, fueled in part by the prospect that it should be a room temperature superconductor \cite{ashcroft_metallic_H, ginzburg_metallic_H}.
In ordinary conditions, bulk hydrogen exists in molecular form, which can be dissociated at sufficiently high temperatures and/or pressures.
Liquid atomic hydrogen has been produced in dynamic compression experiments \cite{liquid_H_pnas, liquid_H_scirep, liquid_H_prl, liquid_H_prb}, but the atomic solid has yet to be unambiguously observed in the lab.

The pressure required to produce the atomic solid is currently estimated to be over 500 GPa.
Experiments have brought the molecular solid above 400 GPa, with no clear signs of dissociation, although metallic behavior is present \cite{H_loubeyre, H_eremets}.
The theoretical estimate has evolved from Wigner and Huntington's initial calculation of 25 GPa \cite{wigner_huntington} to above 570 GPa \cite{sscha_dmc} today.
This latter estimate involved calculating the free energies of different molecular and atomic structures, using advanced techniques to treat electronic correlations and anharmonic lattice contributions as accurately as possible.

Less studied is the temperature at which the solid forms.
The increasingly accurate techniques used to describe lattice vibrations in solids are not applicable to the liquid state.
To treat solid and liquid on equal footing in order to determine their phase boundary, an atomistic simulation such as molecular dynamics (MD) or Monte Carlo must be done.
Additionally, the small mass of hydrogen necessitates the use of path integrals, especially at low temperatures where the solid forms.

Existing studies based on ab-initio, path-integral molecular dynamics have produced mixed results.
Chen et al. found the melting point to be 200 K at 500 GPa, decreasing to nearly 0 K at 1 TPa \cite{aimd_melting_2013}.
Liu et al. found a higher melting point of about 250 K at 500 GPa \cite{aimd_melting_jpcc}.
Geng, Hoffmann, and Wu produced a melting line above 200 K with no sign of decreasing to 0 at 1 TPa \cite{aimd_melting_2015}.
These melting lines are shown in Figure \ref{fig:literature} alongside ours, the subject of the present study.

All of these previous studies used density functional theory (DFT) with the Perdew-Burke-Ernzerhof (PBE) functional \cite{PBE}.
Here, we used reptation quantum Monte Carlo (RQMC) \cite{rqmc}, a more accurate method for calculating electronic ground state properties.
This is important because the choice of density functional significantly affects the structures and phase boundaries of hydrogen \cite{dft1, dft2, dft3, dft4}, and the correct choice is unknown.
Each of these previous studies also takes a unique approach to their MD simulations; for example, each features a different thermostat.
Here, we performed solid and liquid simulations with isobaric, isothermal (NPT) path-integral Monte Carlo.
This ensemble is well suited for locating the phase boundary, and with Monte Carlo the temperature and pressure are imposed in an exact manner.

This coupling of thermodynamic path integrals with RQMC resembles the coupled electron-ion Monte Carlo (CEIMC) method \cite{dewing_thesis, ceimc_carlo}.
Like ab-initio molecular dynamics based on DFT, CEIMC is expensive: an electronic structure calculation must be performed at each step in order to produce the required energies and forces.
This expense was a limiting factor in the previous studies, forcing compromises in system size and choice of ensemble.
We alleviate this bottleneck by using a machine-learned model trained on RQMC calculations, which serves as the surrogate for the RQMC step.

Our model is described in sections \ref{s:model} and \ref{s:data}, including details about our RQMC dataset and its construction \cite{shared_data}.
Section \ref{s:labelling} describes our electronic calculations, while section \ref{s:simulations} describes our atomistic simulations.
The results of these simulations are then given in section \ref{s:results}.

\section{\label{s:methods}Methods}

\subsection{\label{s:model}Model}

We constructed machine-learned interatomic potentials using MACE \cite{mace_design, mace}.
These are equivariant, message-passing neural networks trained on ab-initio energies and forces.
In these models, an inputted configuration of atoms is translated into a set of learnable, equivariant features, which is then mapped to a total energy.
Equivariant models significantly outperform models based on invariant scalar features in terms of overall goodness-of-fit, training speed, and generalization ability, see \cite{painn, nequip} in addition to the above references.

MACE distinguishes itself from other equivariant message-passing networks by incorporating higher body order in the message-passing step.
Specifically, other schemes only involve two-body terms in this step.
MACE includes terms of higher order by adapting the formalism of the atomic cluster expansion \cite{ace}.
This leads to high accuracy with less message-passing iterations, making MACE less expensive in training and evaluation than other schemes.

We trained MACE models on a carefully curated dataset of energies and forces, described in the next section.
In particular, we trained a model on energies and forces calculated using RQMC; hereafter this model will be referred to our QMC-MACE model of atomic hydrogen.
We also trained models on DFT-labelled data, which were used for testing and exploration.
Our QMC-MACE model has a cutoff radius of 3.0 \AA, corresponding to immediate neighborhoods consisting of 75 -- 149 atoms.
Our final model is actually the sum of this QMC-MACE model and a pair-wise potential, meaning that the QMC-MACE model is trained on the difference between the RQMC energy and this pair potential.
This pair potential is energy of two hydrogen atoms in vacuum as a function of the distance between them \cite{H2_energy}, shifted to vanish at the equilibrium bond length.
The purpose of this term is to enforce repulsion when atoms get very close; it is not necessary for this repulsion to accurately describe the atoms in the bulk, since QMC-MACE corrects it.

\subsection{\label{s:data}Training data}

The configurations that make up our dataset were produced with an iterative, exploratory procedure.
At each iteration, beginning with the available data, an intermediate model is trained and used for exploration.
Our goal was to produce a set of configurations which sufficiently samples the region of configuration space in which the melting of atomic hydrogen occurs; this region is not known \emph{a priori}.
The use of intermediate models allows us to explore configuration space relatively quickly and extensively, so that the best possible assessment can be made of our sampling.
This is in contrast to using, for example, AIMD, which is considerably more expensive, limiting the exploration that can be done.
We employed a variety of techniques in an attempt to efficiently realize a complete and diverse dataset.

Our dataset was initialized with trajectories from AIMD and relaxation, as well as a selection of crystalline structures.
Throughout this section, the density of a given structure is specified by the parameter $r_s$, defined by
\begin{equation}
    \label{eq:r_s}
    \frac{4}{3} \pi \left( \frac{r_{s}}{a_0} \right)^3 = \frac{V}{N}
\end{equation}
where $a_0$ is the Bohr radius, $V$ is the volume of the structure, and $N$ is the number of atoms (which is also the number of electrons).
Four AIMD simulations were performed at two different fixed volumes ($r_s = 1.167, 1.2$) and two different fixed temperatures (800 K, 2000 K).
Relaxation calculations were performed on 6 different supercells containing liquids at three different densities ($r_s = 1.1, 1.2, 1.3$).
Every configuration produced through the relaxation calculations is included in the dataset.
We also included supercells of the BCC, FCC, and diamond structures at 50 different densities between $r_s = 1.1$ and $r_s = 1.3$; other ordered structures were produced by intermediate models.

For exploration, the methods used to generate candidate configurations varied between every iteration.
In addition to trajectories produced by simulations in various equilibrium thermodynamic conditions, we also performed simulations in which the temperature or pressure is incrementally changed every few steps.
We also produced two-phase configurations by performing simulations in which only half of the atoms were moved.
Beginning with a crystalline configuration, all of the atoms in one spatial half of the cell are fixed, and a simulation is carried out at high temperature, resulting in configurations which are half liquid and half solid.
These configurations contain unique local environments at the interface for models to learn from.

Our DFT-labelled dataset totals 9982 configurations, covering densities from $r_s = 1.11$ to $r_s = 1.29$, corresponding to pressures between 400 and 1000 GPa.
This set features a variety of supercell shapes and sizes, with the number of atoms varying between $N = 100$ and $N = 256$.
A model trained on this dataset is able to reproduce the 350 K classical melting line observed in \cite{aimd_melting_2015}.
This model reproduces previous structure searching studies \cite{atomic_H_airss, fddd}, finding the expected Cs-IV structure (space group $I 4_1 / a m d$), but also finding a distortion ($F d d d$) and a molecular structure ($C m c a$) at 500 GPa nearby in enthalpy.
The liquid structure seen in AIMD simulations \cite{old_liquid_hydrogen} is also reproduced by this model.
In conjunction with our exhaustive exploration, these ``real-world'' tests indicated to us that the dataset contained the necessary information to model melting and the nearby solid and liquid phases.

Using farthest point sampling \cite{supp} we selected a subset of 2000 of the 9982 configurations for labelling with RQMC.
To test that this subset was adequate for our purposes, we simulated statistical noise on the DFT energies and forces.
We found that a model trained on this noisy subset had approximately the same accuracy as a model trained on the full dataset.
This test model also produces the same classical melting behavior as the full model, within our resolution.
The root mean square fitting error of the forces in our QMC-MACE model is 0.120 eV/\AA, consistent with what we expected from our simulated noise test.
Based on our noise test, we believe that the true fitting error to be closer to 0.08 eV/\AA, which is inflated by the noise present in the RQMC data.
The RMS fitting error of the energies is 0.03 eV/atom, which is higher than what the noise can account for.
We believe this is because the QMC force estimator is not exactly consistent with the derivative of the energy.
For more on training with QMC data, see \cite{qmc_training}.

\subsection{\label{s:labelling}RQMC and DFT}

In RQMC, the ground state energy of an electronic hamiltonian $H$ is obtained by the projection $e^{-\tau H} \ket{\psi}$ on a trial wavefunction $\psi$.
For an observable $A$ which is diagonal in real space $x$, the ground state expectation may be estimated with the path integral
\begin{equation}
    \label{eq:rqmc_observable}
    \mean{A} \approx \frac{
        \int dx_0 \dots dx_{2M} A(x_M) \psi(x_0) \psi(x_{2M}) \prod_{i = 1}^{2M} \mean{x_{i-1} | e^{-\Delta \tau H} | x_i}
    }{
        \int dx_0 \dots dx_{2M} \psi(x_0) \psi(x_{2M}) \prod_{i = 1}^{2M} \mean{x_{i-1} | e^{-\Delta \tau H} | x_i}
    },
\end{equation}
where $M$ is the number of divisions of the operator $e^{-\tau H}$, so that $\Delta \tau \equiv \tau / M$.
This is applicable to the force estimator of \cite{ccz}, which we used for this work.
To calculate the total energy, the same distribution holds, but the local energy must be calculated at the ends of the path $x_0, x_{2M}$ rather than at the middle $x_M$.
In the reptation algorithm \cite{rqmc, bounce}, this distribution is sampled by having the path slither like a reptile: the path is moved by removing one end (the ``tail'') and adding to the other end (the ``head'').

RQMC is related to diffusion Monte Carlo (DMC) \cite{metropolis_ulam, qmc_rmp} in that they both perform the projection $e^{-\tau H}$.
In DMC, this projection is mapped onto a diffusion problem.
For a real-space observable $A$ (such as our force estimator), DMC calculates the mixed estimate $\mean{\psi | e^{-\tau H} A | \psi}$, in which the projection only occurs on one side, whereas the ``pure'' estimate is \eqref{eq:rqmc_observable}.
Our energy and force calculations are thus comparable to or slightly better than what would be achieved with DMC.

Trial wavefunction optimization and subsequent RQMC calculations were performed with \texttt{QMCPACK} \cite{qmcpack_2018, qmcpack_2020}.
We used trial wavefunctions of the single Slater-Jastrow form, where the orbitals forming the determinant are produced by DFT.
The Jastrow factor was individually optimized for every configuration in the dataset.
All projections used a reptile consisting of 200 beads, and a timestep of 0.02 Ha, corresponding to a total projection time of $\tau = 4$ Ha.
Twist averaging was performed over either a shifted $7 \times 7 \times 7$ or $6 \times 6 \times 6$ grid; supercells containing 200 or more atoms used the latter grid.
Aside from the use of supercells and twist averaging, no additional finite-size corrections \cite{CCMH} are added.
The reason for this is that our model formally requires that the energies and forces be consistent, but these additional corrections are only applicable to the energy.
The fixed-node approximation is required, which is enforced by rejecting moves that change the sign of $\psi(x_0) \psi(x_{2M})$.

DFT calculations were performed with \texttt{Quantum ESPRESSO} \cite{QE_2009, QE_2017}.
A Troullier-Martins norm-conserving pseudopotential with a radial cutoff of 0.65 bohr is used to represent the hydrogen cores.
A plane-wave kinetic energy cutoff of 150 Ry was used, and Brilluoin zone integration was done over a shifted $7 \times 7 \times 7$ grid.
For labelling our dataset, we used the PBE functional, which allowed us to compare DFT-trained models with existing literature.
When generating orbitals for RQMC, we used the PZ LDA functional \cite{PZ}, which resulted in slightly improved variances.

\subsection{\label{s:simulations}Simulations}

We determined melting temperatures by performing two-phase simulations.
The system is initialized as a 784-atom supercell which is half crystalline, half liquid.
We then do a simulation at fixed pressure and fixed temperature, and the system either becomes entirely liquid or entirely crystalline.
By repeating this with different pressures and temperatures, we can narrow down the melting points.

The crystalline half is represented by the Cs-IV structure ($I4_1/amd$), the primary candidate for solid atomic hydrogen up to terapascal pressures \cite{atomic_H_airss, sscha_dmc}.
A distortion of this structure ($Fddd$) has previously been proposed on the basis of its appearance in structure searches \cite{fddd, aimd_melting_2015}, but we found that this distortion does not survive thermal fluctuations, so we do not consider this structure here.
Cs-IV is a tetragonal structure with only two free parameters, the $c/a$ ratio and the density.
These two parameters are determined by performing PIMC on a fully crystalline supercell.

The liquid half is constructed by performing classical simulations at fixed temperature and fixed volume (NVT).
The volume is fixed to match the shape of the corresponding crystalline half, so that the two cells can easily be combined.
We performed classical rather than quantum simulations for this because we found that they result in similar liquid structures, with the primary difference being that the quantum simulations resemble the classical ones at much higher temperatures.

We also performed single-phase simulations to calculate the volume and energy at different temperatures and pressures.
These are then used to estimate the change in volume $\Delta V$ and energy $\Delta U$ upon melting, as well as the latent heat of fusion $\lambda \equiv \Delta U + P \Delta V$.
These changes can be used to estimate the derivative of the melting line via the Clausius-Clapeyron relation \cite{fermi}
\begin{equation}
    \label{eq:clausius_clapeyron}
    \frac{dP}{dT} = \frac{\lambda}{T \Delta V}.
\end{equation}
We used these estimates to further refine our determination of the melting lines.

In Monte Carlo simulations at fixed temperature and pressure, particles are moved $R \to R'$ according to a rule $T(R \to R')$ and then accepted with probability
\begin{equation}
    \label{eq:acceptance_ratio}
    A(R \to R') = \frac{\Pi(R') T(R' \to R)}{\Pi(R) T(R \to R')}
\end{equation}
where $\Pi$ is the target distribution.
In the NPT ensemble, the appropriate distribution is \cite{allen_tildesley}
\begin{equation}
    \label{eq:distribution}
    \Pi \propto e^{-\beta (U + P V) + N \log V}
\end{equation}
where $\beta \equiv (k T)^{-1}$ is the inverse temperature, $U$ is the energy, $P$ is the pressure, $V$ is the volume, and $N$ is the number of atoms.

We alternate between two moves: one in which all of the particles are moved while the cell remains fixed, and one in which the particle's reduced coordinates remain fixed while the the cell is changed.
The particle move is
\begin{equation}
    \label{eq:particle_move}
    \begin{split}
        T(\{\boldsymbol{R}\} \to \{\boldsymbol{R}\}') &\propto \prod_i \exp \left( - \frac{\boldsymbol{R}_i' - \boldsymbol{R}_i - \tau \beta \boldsymbol{F}_i)^2}{4 \tau} \right) \\
        &= \exp \left( -\sum_i \frac{(\boldsymbol{R}_i' - \boldsymbol{R}_i - \tau \beta \boldsymbol{F}_i)^2}{4 \tau} \right)
    \end{split},
\end{equation}
in which each particle's random displacement is normally distributed with $\sigma = \sqrt{2 \tau}$, while also being pushed by the force.
This move resembles the ``smart Monte Carlo'' scheme suggested by \cite{smart_mc}.
Note that we move all particles simultaneously, as our model only ever calculates the total energy and all forces at once, just like an electronic structure method would.
The cell move is
\begin{equation}
    \label{eq:cell_move}
    T(\boldsymbol{c} \to \boldsymbol{c}') \propto
    \begin{cases}
        1, \quad & \boldsymbol{c}' - \boldsymbol{c} \in [-\sigma_l, \sigma_l)^3 \times [-\sigma_a, \sigma_a)^3 \\
        0, \quad & \text{otherwise},
    \end{cases}
\end{equation}
where $\boldsymbol{c} \equiv (a, b, c, \alpha, \beta, \gamma)$ are the three side lengths and three angles of the cell.
The random change in each cell component is uniformly distributed within $[\sigma, \sigma)$, where this ``length'' $\sigma$ can be separately chosen for the side lengths and the angles.
In all of our simulations we found distortions of the orthorhombic cell to be irrelevant, so that $\sigma_a = 0$ for all results shown here.
The step sizes $\tau$ and $\sigma$ are chosen such that the acceptance rates are 20\% -- 50\%.

To do path-integral simulations, the physical system is replaced by $p$ copies of the system, or beads \cite{allen_tildesley, helium, tuckerman, feynman_hibbs}.
Using the primitive approximation, this extended system follows the distribution \eqref{eq:distribution} but with an effective potential energy
\begin{equation}
    \label{eq:effective_potential}
    \tilde{U} = \frac{p m}{2 \beta^2 \hbar^2} \sum_{i=1}^p \sum_{j = 1}^N |\boldsymbol{R}_j^i - \boldsymbol{R}_j^{i+1}|^2 + \frac{1}{p} \sum_{i=1}^p U(\{\boldsymbol{R}\}_i),
\end{equation}
where $m$ is the mass of the particles, $\{\boldsymbol{R}\}_i$ is the $i$th bead, and $\boldsymbol{R}_{j}^i$ is the coordinate of particle $j$ in bead $i$.
The moves \eqref{eq:particle_move} and \eqref{eq:cell_move} can still be applied.
The simulation of different isotopes, such as deuterium, can be carried out using the same model $U$ but a different mass $m$, which affects the first term in \eqref{eq:effective_potential}.
We used paths with 10 beads for melting simulations, and 16 beads for equilibrium energy and volume estimates.

\section{\label{s:results}Results}

\begin{figure}
    \begin{subfigure}{0.49\linewidth}
        \includegraphics[width=\textwidth]{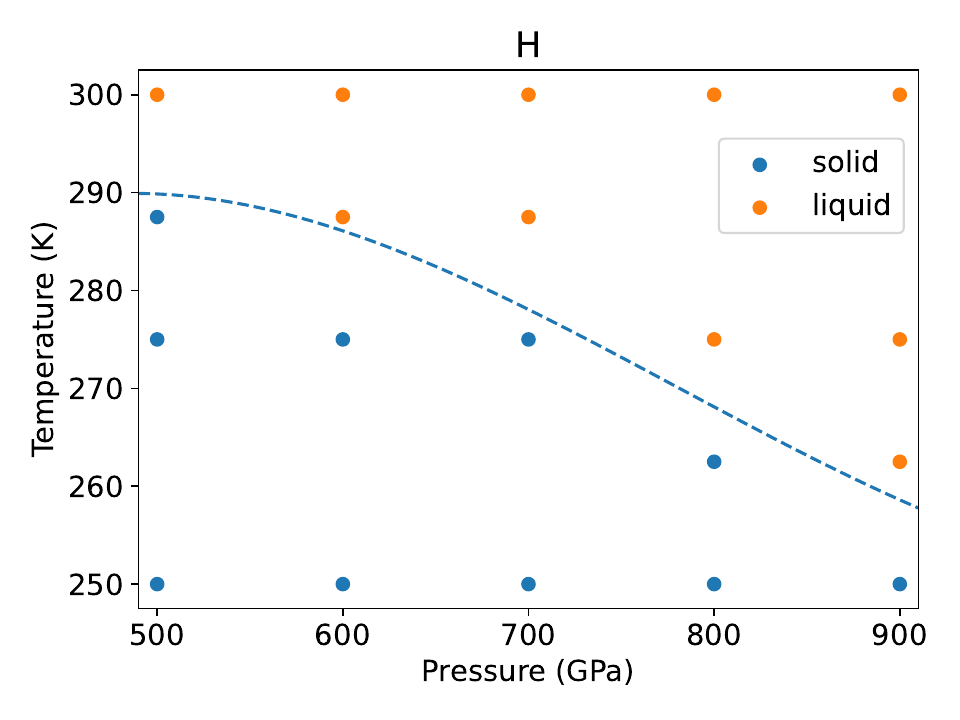}
        \caption{\label{fig:H_phases}}
    \end{subfigure}
    \begin{subfigure}{0.49\linewidth}
        \includegraphics[width=\textwidth]{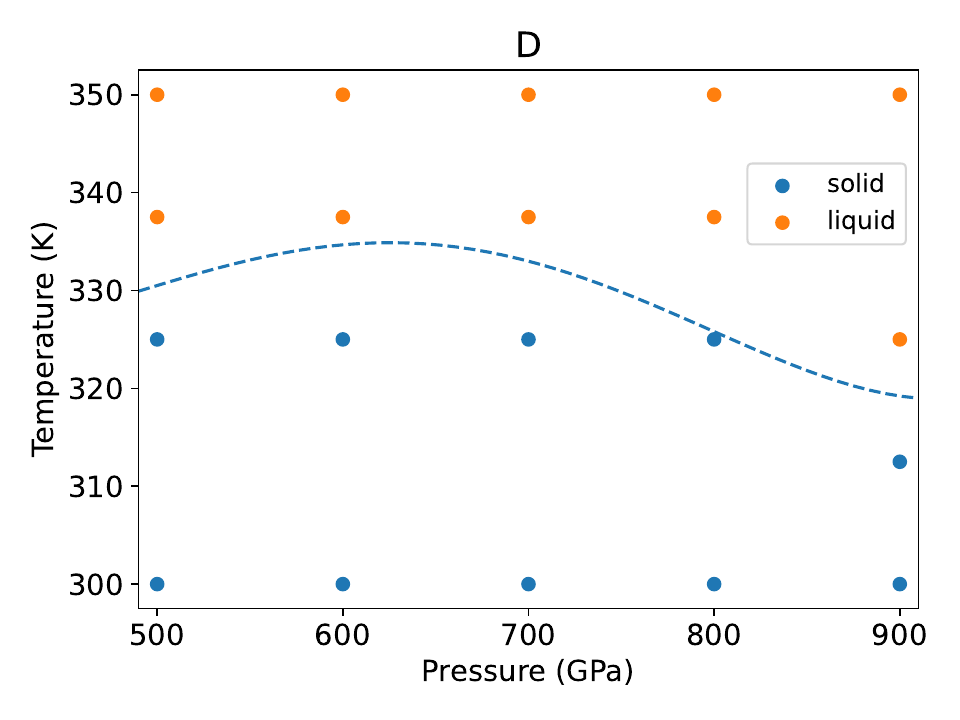}
        \caption{\label{fig:D_phases}}
    \end{subfigure}
    \caption{\label{fig:phases}
        Outcomes of our two-phase simulations for (a) hydrogen and (b) deuterium.
        Each simulation results in either solid or liquid, depending on the chosen temperature and pressure.
    }
\end{figure}

Shown in Figure \ref{fig:phases} are the results of our two-phase simulations for both hydrogen and deuterium.
Each point represents a simulation at a particular temperature and pressure, and the color indicates the outcome.
Our estimated melting lines are the dashed lines drawn between the phases; the shapes of these lines are discussed below.
For hydrogen, the melting line slightly but clearly decreases as pressure is increased.
Quantum effects decrease the melting temperature, as the more massive deuterium has a higher melting line, and the classical line is even higher.

\begin{figure}
    \begin{subfigure}{0.49\linewidth}
        \includegraphics[width=\textwidth]{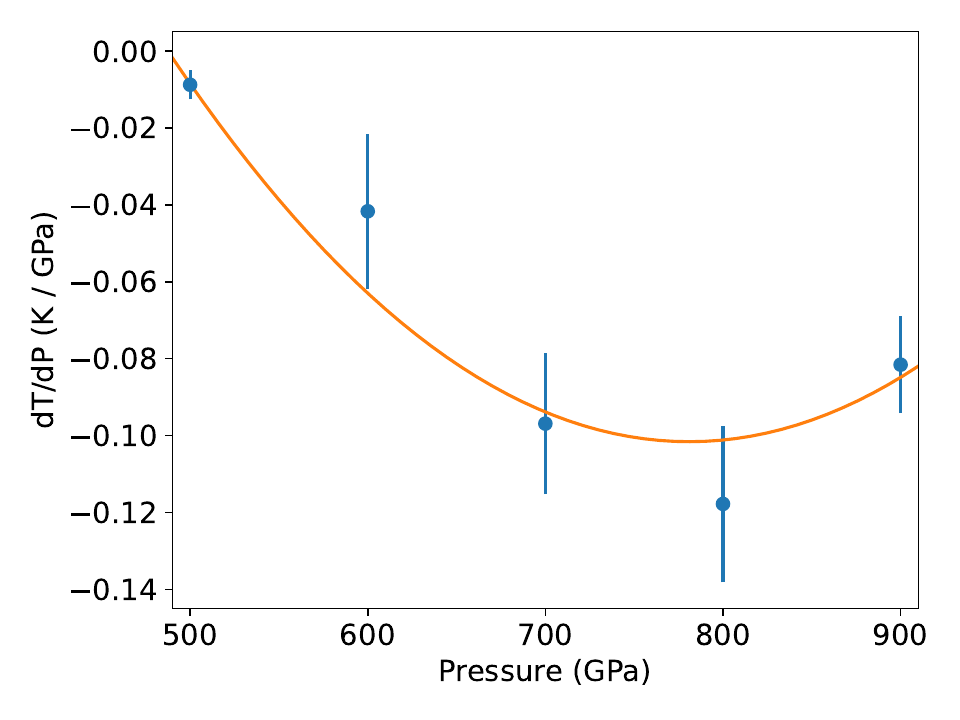}
        \caption{\label{fig:derivative}}
    \end{subfigure}
    \begin{subfigure}{0.49\linewidth}
        \includegraphics[width=\textwidth]{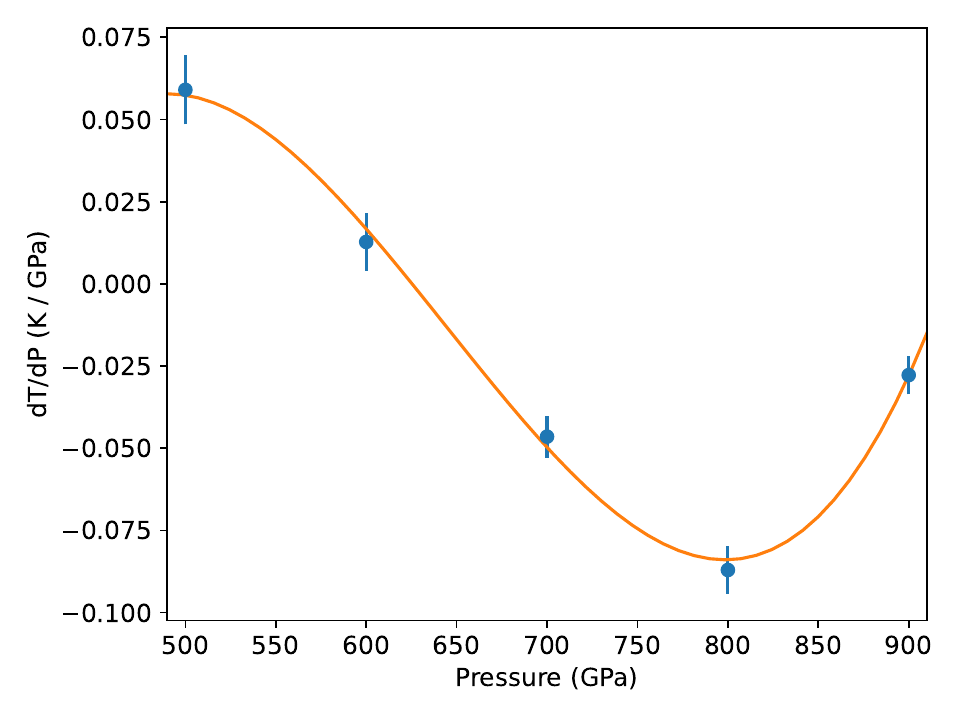}
        \caption{\label{fig:D_derivative}}
    \end{subfigure}
    \caption{\label{fig:derivatives}
        The derivative $dT/dP$ of the melting line for (a) hydrogen and (b) deuterium.
        The Clausius-Clapeyron estimates are given in blue, and the derivatives of the lines in Figure \ref{fig:phases} are in orange.
    }
\end{figure}

The results of our single-phase simulations are summarized in Table \ref{tab:numbers}.
These are used to estimate the derivative of the melting line at various pressures with \eqref{eq:clausius_clapeyron}.
Each melting line is represented using a polynomial which simultaneously fits these derivatives as well as the solid-liquid bounds from Figure \ref{fig:phases}.
Shown in Figure \ref{fig:derivatives} are the derivatives of the melting lines estimated in this manner (blue), alongside the derivatives of the lines drawn in Figure \ref{fig:phases}; for more convenient comparison, the reciprocal of \eqref{eq:clausius_clapeyron} is plotted.
We can clearly resolve a decrease in the melting line in hydrogen.
The sign of $dT/dP$ is given by the sign of $\Delta V$, since the change in entropy (or latent heat) from solid to liquid is positive, as we would expect.
For hydrogen, the volume shrinks upon melting throughout this pressure range, albeit by a small amount, which is also consistent with the flatness of the melting line.
The error bars shown in these figures are estimated with a jackknife procedure \cite{jackknife, supp}.

\begin{table*}
    \caption{\label{tab:numbers}
        Melting temperatures, volume and energy changes, as estimated with PIMC using our QMC-MACE model.
        Since the two-phase simulations produce upper and lower bounds to the melting point, the temperatures are given here as intervals.
        A negative volume change indicates contraction upon melting, i.e. the liquid is more dense.
    }
    \begin{ruledtabular}
        \begin{tabular}{lccccc}
            Isotope & $P$ (GPa) & $T_m$ (K) & $\Delta V$ (\AA$^3$/atom) & $\Delta U$ (eV/atom) \\
            \colrule
            H & 500 & 287.5--300 & -0.00013(5) & 0.029(3) \\
            H & 600 & 275--287.5 & -0.0007(3) & 0.032(2) \\
            H & 700 & 275--287.5 & -0.0015(2) & 0.034(2) \\
            H & 800 & 262.5--275 & -0.0019(2) & 0.037(2) \\
            H & 900 & 250--262.5 & -0.00127(5) & 0.032(3) \\
            D & 500 & 325--337.5 & 0.0010(2) & 0.033(2) \\
            D & 600 & 325--337.5 & 0.0002(1) & 0.0360(6) \\
            D & 700 & 325--337.5 & -0.0008(1) & 0.040(1) \\
            D & 800 & 325--337.5 & -0.0015(1) & 0.0427(5) \\
            D & 900 & 312.5--325 & -0.0005(1) & 0.0369(5) \\
        \end{tabular}
    \end{ruledtabular}
\end{table*}

\begin{figure}
    \begin{subfigure}{0.49\linewidth}
        \includegraphics[width=\textwidth]{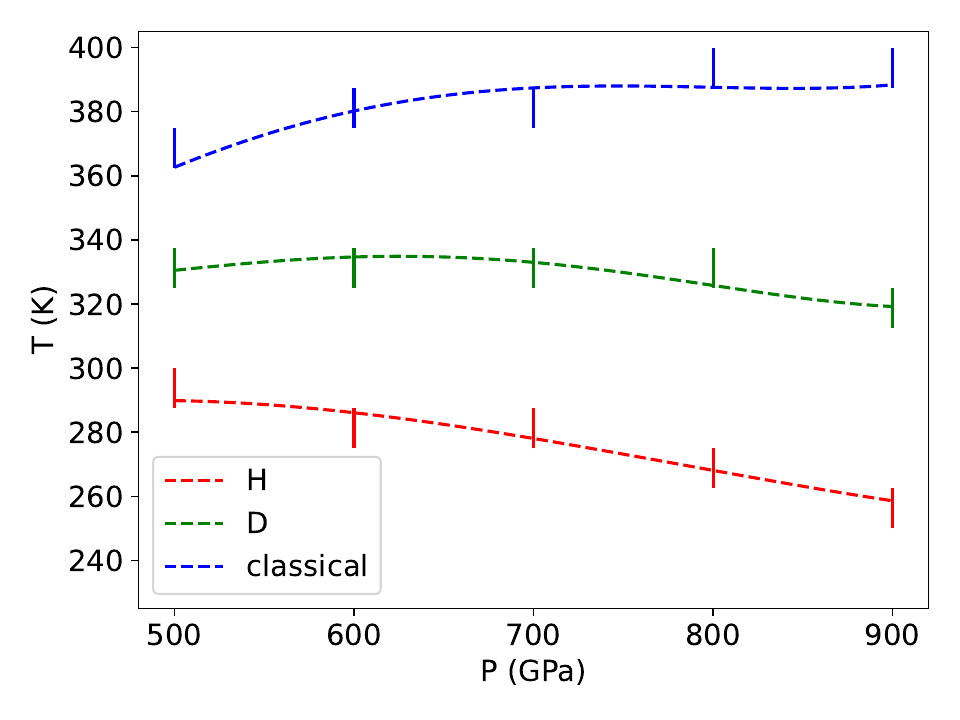}
        \caption{\label{fig:quantum_effects}
        }
    \end{subfigure}
    \begin{subfigure}{0.49\linewidth}
        \includegraphics[width=\textwidth]{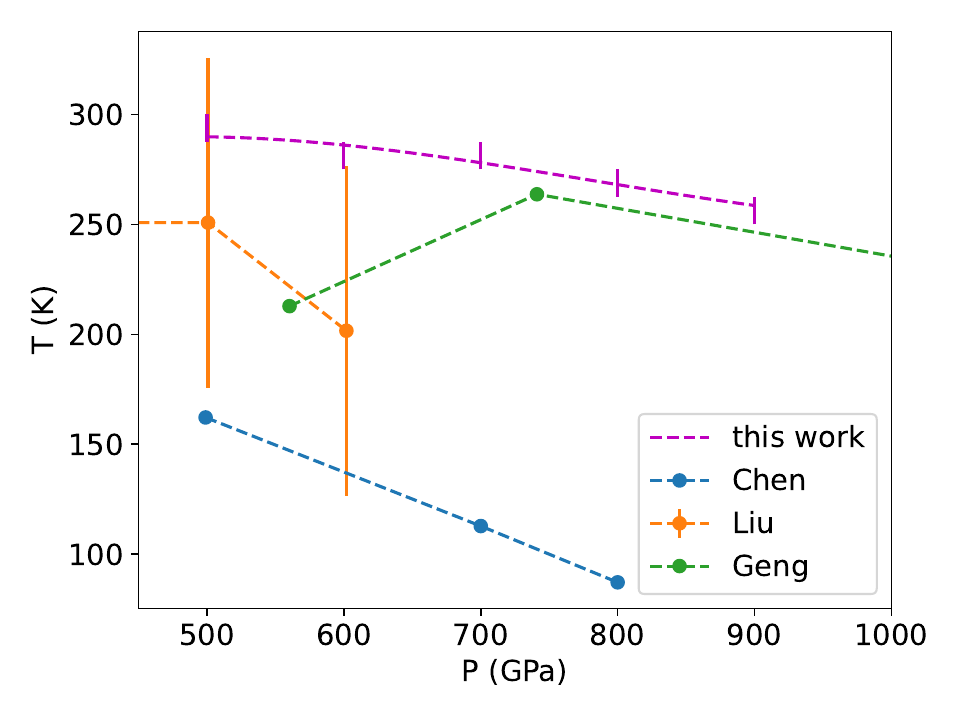}
        \caption{\label{fig:literature}
        }
    \end{subfigure}
    \caption{\label{fig:summary}
            (a) The melting lines of hydrogen and deuterium, as well as the melting line from classical simulations.
            All three results use the same QMC-MACE model.
            The error bars shown here are the upper and lower bounds that can be determined from Figures \ref{fig:H_phases} and \ref{fig:D_phases}.
            (b) Our melting line for hydrogen compared with previous results.
    }
\end{figure}

To illustrate quantum effects, we simultaneously plot all of our melting lines in Figure \ref{fig:quantum_effects}.
The lines for H and D are the same as in Figure \ref{fig:phases}, and the classical melting line was calculated in a manner similar to that of H and D but without the use of path integrals \cite{supp}.
The inclusion of quantum effects clearly reduces the melting line by about 100 K, which has also been observed in previous studies \cite{aimd_melting_2013, aimd_melting_jpcc, aimd_melting_2015}.
Additionally, we can resolve a slight change in the slope of the melting line, which is slightly positive in the classical case.
The classical system expands upon melting, while the quantum system shrinks.

Shown in Figure \ref{fig:literature} is our result for hydrogen compared with previous work: \cite{aimd_melting_2013} in blue, \cite{aimd_melting_jpcc} in orange, and \cite{aimd_melting_2015} in green.
Our melting line is higher than all other previous results, indicating that the solid is stable to slightly higher temperatures.
Although there are many factors contributing to the scatter among all of these results, we attribute our higher melting line to the use of RQMC (all other lines feature PBE DFT) and finite-size effects.
We tested these two factors independently using classical two-phase melting simulations and found that both lead to an increase in the melting temperature \cite{supp}.

\section{\label{s:conclusions}Conclusions}

Using path-integral Monte Carlo, we simulated the melting of atomic hydrogen and deuterium.
We found that the melting line of atomic hydrogen slightly but clearly decreases with pressure up to 900 GPa.
In addition to significantly lowering the melting temperature, we found that quantum effects change the sign of the melting line's slope.
Previously simulated melting lines were lower than ours, and were inconclusive regarding the slope of the melting line up to 1 TPa.
Our two-phase simulations feature the isobaric-isothermal ensemble, larger supercells, and more accurate energies and forces with RQMC.
This was made possible through the use of a machine-learned model of the energies and forces, which we believe to be an accurate representation of the underlying electronic structure method.

Our higher melting line is still below the estimated superconducting temperature of over 300 K \cite{hydrogen_tc_harmonic, hydrogen_tc_sscha}.
The presence of superconductivity is not considered in our study, as all energy and force calculations here assume the system to be in a normal, metallic state.
It is not clear to us how the presence of a superconducting state would affect the relative stabilities of the solid and liquid.
The free energy of the superconducting solid would be lower than what is simulated here, which would increase the melting point.
However, in \cite{liquid_superconductor}, it was proposed that the liquid state could still host superconductivity, in which case the effect on the melting point would be unclear.

The decreasing melting line lends itself to the intriguing possibility of a superfluid state for the protons.
This is what we might expect if the liquid is stable to sufficiently low temperatures, as quantum effects are clearly strong for atomic hydrogen.
Unlike liquid helium, the electronic state of liquid atomic hydrogen is metallic, or indeed even superconducting.
The confluence of superconductivity and superfluidity has been considered in \cite{superH_2004, superH_2005}.
However, the decrease in the melting line may not continue indefinitely. 
For example, QMC calculations at much higher densities suggest a melting point of 1680 K at 24 TPa \cite{very_high_pressure}.

The possibility of protonic superfluidity is excluded in our study since we treat the protons as distinguishable particles in our path integrals.
Even without superfluidity (which we do not expect at these temperatures), neglecting the fermionic character of the protons is still erroneous.
However, in the conditions studied here, we estimate that this error in the free energy is on the order of 1 K, based on fermionic PIMC using a pair potential adjusted to have the same melting temperature.

\section{\label{sec:acknowledgements}Acknowledgements}

We would like to thank Markus Holzmann and Carlo Pierleoni for their insights and dicussions on atomic hydrogen and QMC.
We would also like to thank Miguel Morales, as we built upon some of his (unpublished) DFT-PIMD work.
K. L. would like to thank Yubo Yang, Raymond Clay, and Jeffrey McMahon for their assistance on various practical aspects of this work.
This work was supported by the U.S. Department of Energy, Office of Science, Office of Basic Energy Sciences, Computational Materials Sciences program under Award No. DE-SC0020177.
This work utilized computing resources from: Summit, operated by the Oak Ridge Leadership Computing Facility, a DOE Office of Science User Facility supported under Contract No. DE-AC05-00OR22725; Perlmutter, operated by the National Energy Research Scientific Computing Center (NERSC), supported under Contract No. DE-AC02-05CH11231 using NERSC award ALCC-ERCAP0029712; HAL, operated by the National Center for Supercomputing Applications, supported by the NSF’s Major Research Instrumentation Program, Grant No. 1725729, as well as the University of Illinois at Urbana-Champaign.
\bibliography{bibliography.bib}

\end{document}


\preprint{}

\title{SM: Melting of solid atomic hydrogen with path-integral Monte Carlo}


\author{Kevin K. Ly}

\author{David M. Ceperley}
\affiliation{Department of Physics, University of Illinois at Urbana-Champaign}


\date{\today}


\maketitle


\section{Farthest point sampling}

Our dataset was constructed using an iterative, exploratory procedure.
At each iteration, a pool of candidate configurations is produced using an intermediate model, and a subset of these candidates is selected for labelling and addition to the existing dataset.
Every configuration in the current dataset and the candidate set is transformed into a set of features $\{\boldsymbol{R}\} \to \boldsymbol{x}$.
Candidates are then selected using farthest point sampling, whereby points are added to the dataset one-by-one based on their distance to the current dataset.
Let $X$ be the current dataset, and let $Y$ be a set of candidates.
For each candidate $\boldsymbol{y}_i \in Y$, consider the distance to the nearest neighbor in X
\begin{equation}
    \label{eq:nearest_neighbor_distance}
    d_i \equiv \min_{\boldsymbol{x} \in X} |\boldsymbol{y}_i - \boldsymbol{x}|
\end{equation}
where $|\cdot|$ is the $l^2$ norm.
We then select the candidate whose distance to the dataset is greatest, and add it to the set $X$.
The process is then repeated until a satisfactory number of selections have been made.
Note that the nearest neighbor distance \eqref{eq:nearest_neighbor_distance} generally changes throughout this process, since $X$ is being updated as selections are made.

\begin{figure}
    \begin{subfigure}{0.49\linewidth}
        \includegraphics[width=\textwidth]{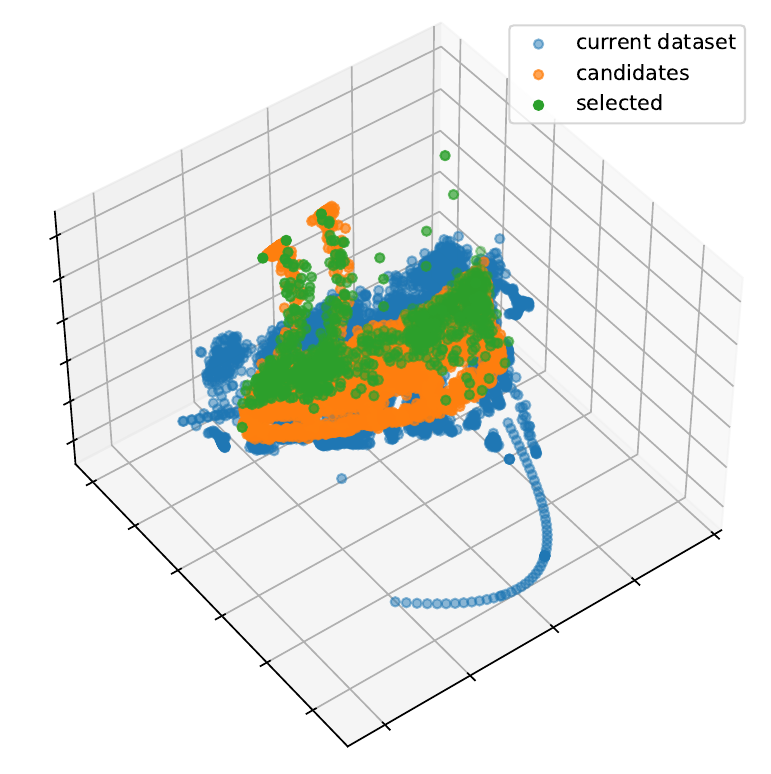}
        \caption{\label{fig:fps_example}}
    \end{subfigure}
    \begin{subfigure}{0.49\linewidth}
        \includegraphics[width=\textwidth]{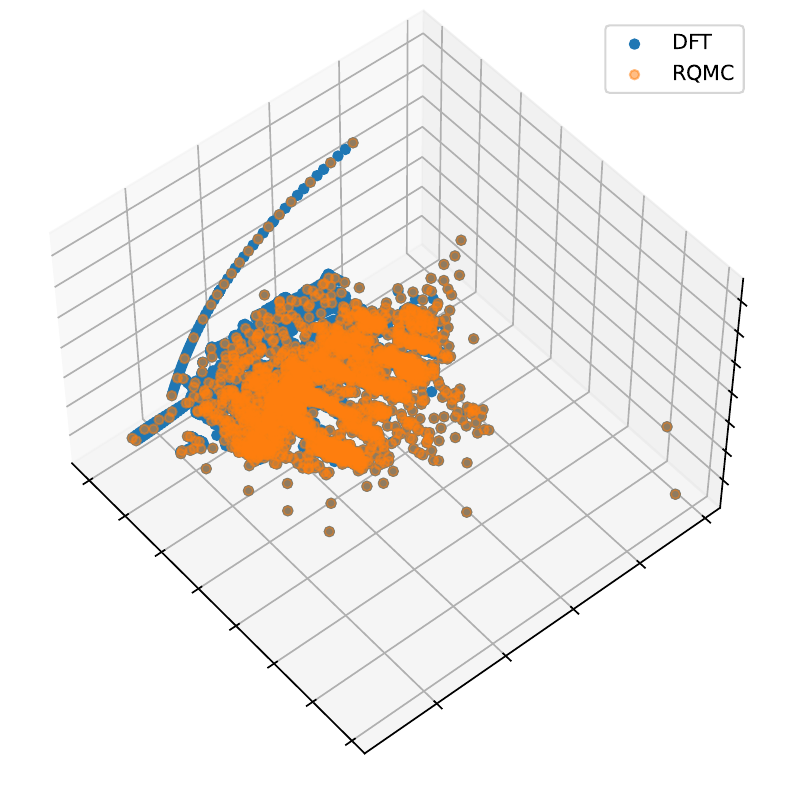}
        \caption{\label{fig:dataset}}
    \end{subfigure}
    \caption{\label{fig:model}
        (a) Example of farthest point sampling, in a SOAP feature space of reduced dimension.
        Each point represents a different configuration, and the axes have arbitrary units.
        Points (configurations) are selected from a candidate set based on their distance to the current set.
        (b) Full dataset of DFT-labelled configurations (blue) and the subset of points chosen for labelling with RQMC (orange).
    }
\end{figure}

An example of this selection process is shown in Figure \ref{fig:fps_example}.
Every configuration in the existing and candidate configurations is represented by a vector of SOAP features \cite{soap, asap}.
Each point in the scatterplot represents a configuration in this SOAP space projected onto the first three principal components; note that we perform selection in the full space, and the dimensionality reduction is only used here for visualization.
In this example, there are 8822 existing configurations (blue), 8017 candidate configurations generated by an intermediate model (orange), and we selected selected 1000 of these candidates (green) for labelling.
Farthest point sampling ensures that we choose candidates which have the least overlap with existing configurations, minimizing redundancy.

We also used farthest point sampling to select a subset of the structures of labelling with RQMC.
The full set of structures is shown in Figure \ref{fig:dataset}, and the selected structures are colored orange.

\section{Classical melting}

\begin{figure}
    \begin{subfigure}{0.49\linewidth}
        \includegraphics[width=\textwidth]{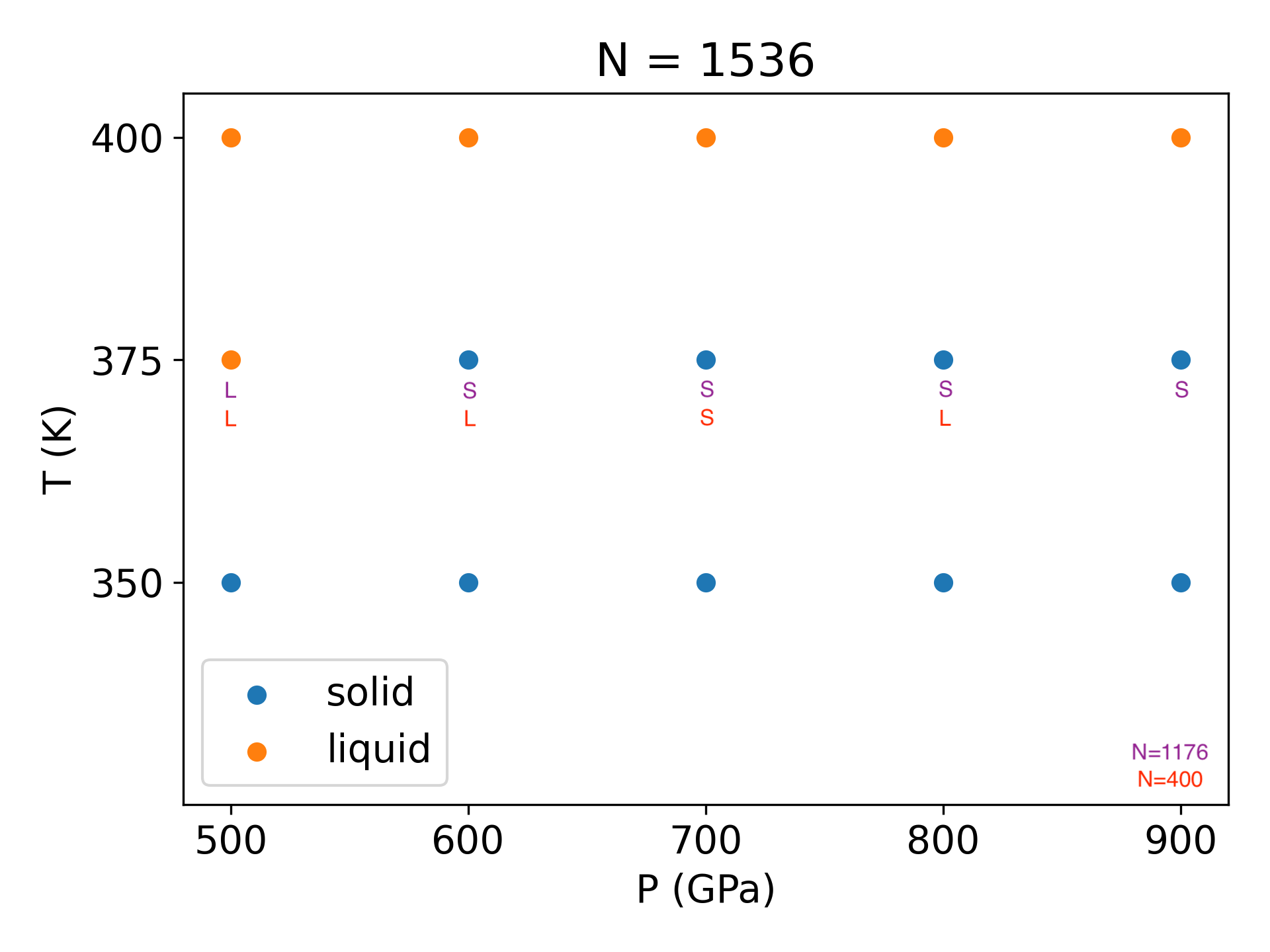}
        \caption{\label{fig:finite_size}}
    \end{subfigure}
    \begin{subfigure}{0.49\linewidth}
        \includegraphics[width=\textwidth]{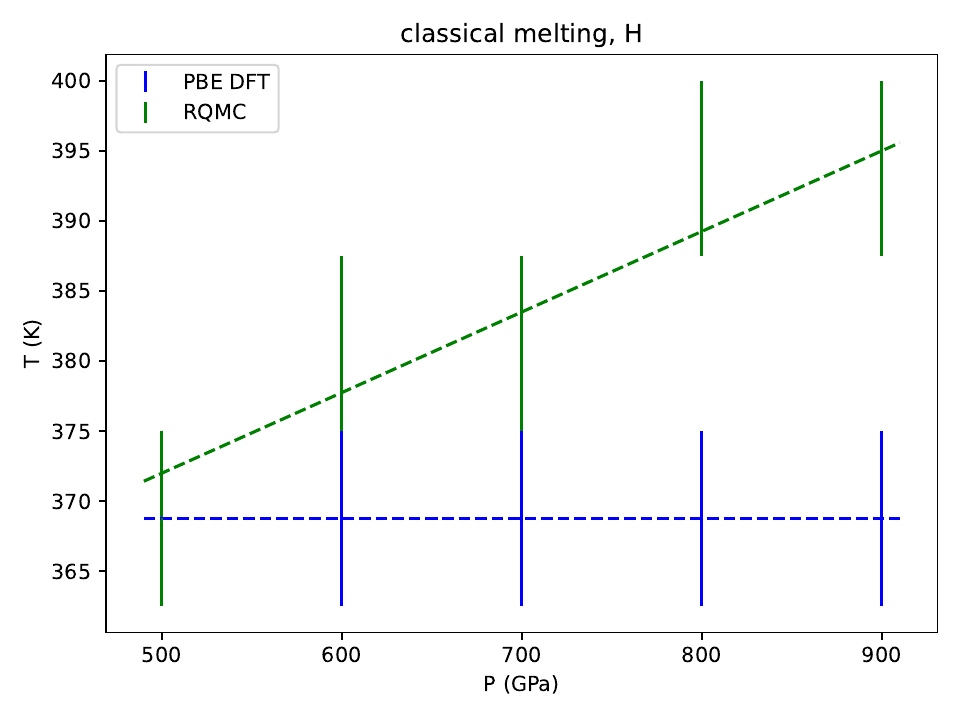}
        \caption{\label{fig:dft_qmc}}
    \end{subfigure}
\end{figure}

Using our RQMC- and DFT-trained models, we performed classical melting simulations.
These feature a similar two-phase setup to the path-integral simulations shown in the main text, but without the use of path integrals.

Shown in Figure \ref{fig:finite_size} are classical melting simulations with various system sizes, using our QMC-MACE model.
The colored points on the figure represent the outcomes of simulations using $N = 1536$.
Additional simulations were performed using $N = 400, 1176$ at $T = 375$ K, and the results of these are indicated just below each point at this temperature.
In these additional annotations, ``S'' means that the outcome is solid, and ``L'' means that the outcome of liquid, and the two different colors correspond to the two different sizes.
Simulations with $N = 1176$ match those of $N = 1536$, while the smaller simulations incorrectly melt.
We see also that the smaller simulation has a non-monotonic melting line.

Shown in Figure \ref{fig:dft_qmc} are classical melting lines using our QMC-MACE model (green) and using a model trained on our full DFT dataset (blue).

\section{Distinguishing liquid from solid}

\begin{figure}
    \begin{subfigure}{0.49\linewidth}
        \includegraphics[width=\textwidth]{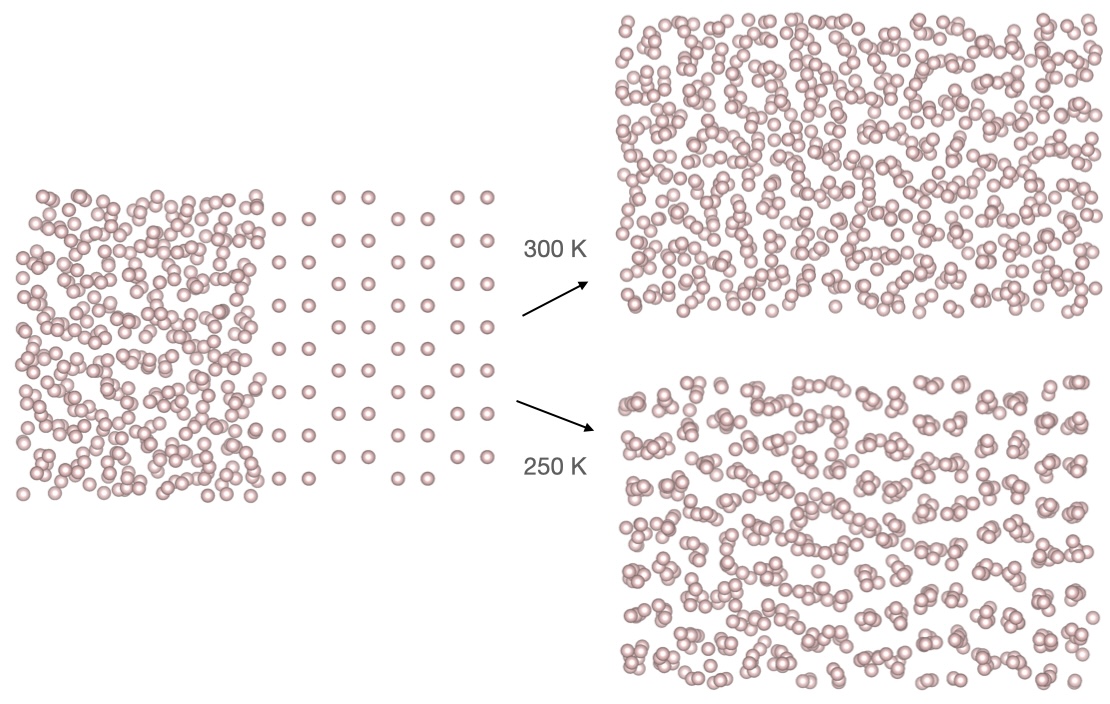}
        \caption{\label{fig:two_phase_pic}}
    \end{subfigure}
    \begin{subfigure}{0.49\linewidth}
        \includegraphics[width=\textwidth]{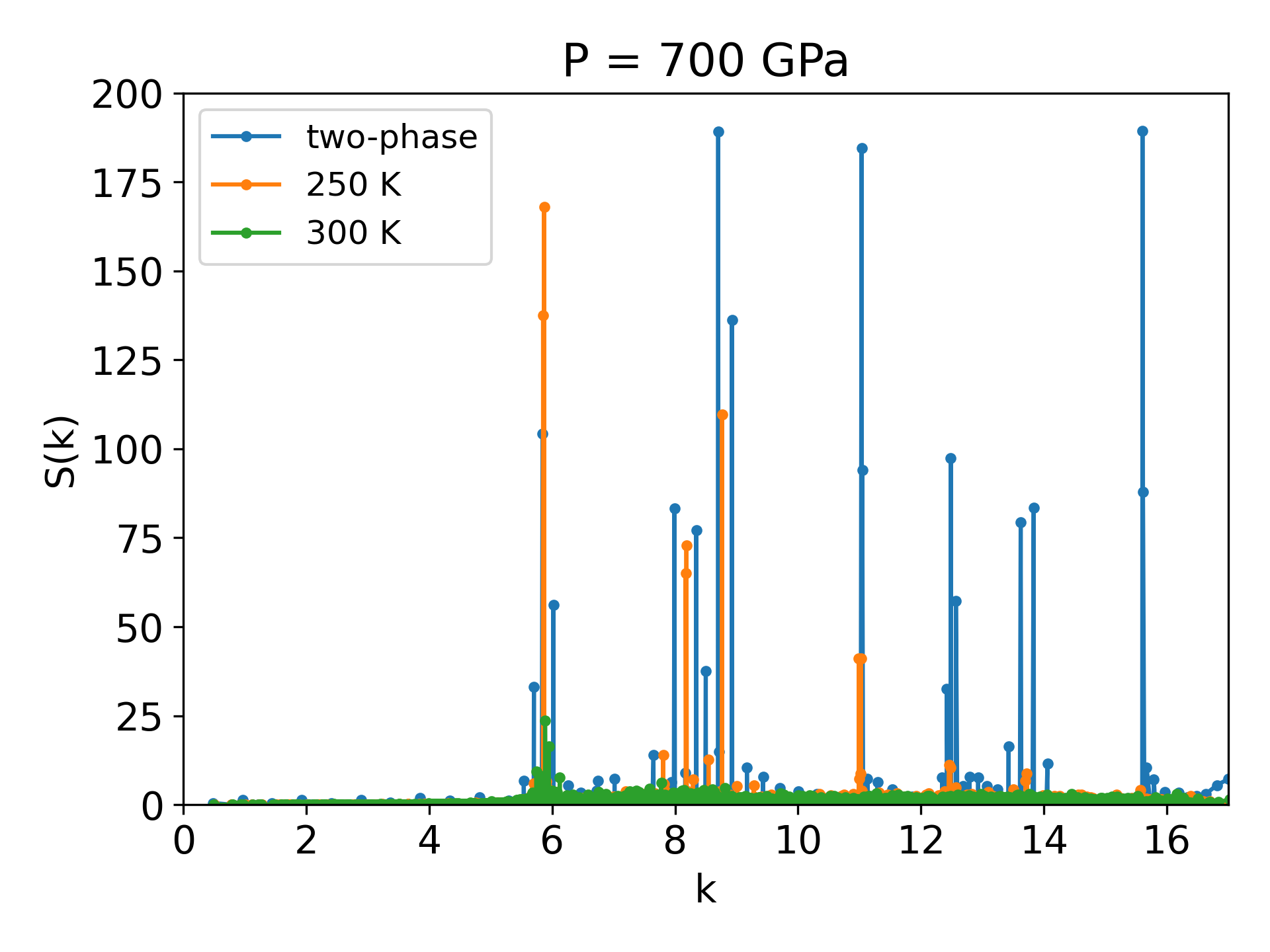}
        \caption{\label{fig:sk_example}}
    \end{subfigure}
    \caption{\label{fig:two_phase_setup}
        (a) An initial liquid-solid sample (left) which transforms into a liquid at 300 K (right, top) or a solid at 250 K (right, bottom).
        Each of these samples is taken from our simulations at 700 GPa.
        (b) Instantaneous static structure factors for each of the three samples shown in (a).
    }
\end{figure}

Shown in Figure \ref{fig:two_phase_pic} is an example of an initial two-phase configuration (left) which resulted in a liquid at 300 K and a solid at 250 K.
Due to large fluctuations, it can be difficult to distinguish the phases merely by inspecting these configurations.
To determine the phase, we instead inspect an \emph{instantaneous} measurement of the static structure factor.
We calculate the structure factor for only a single sample, rather than averaging over an equilibrium trajectory.
An example of this is shown in Figure \ref{fig:sk_example}, which displays the structure factor for each of the three configurations in Figure \ref{fig:two_phase_pic}.
Although the statistical noise is very apparent, the phases can still be distinguished by examining the behavior of the primary peaks.
To distinguish the solid from the initial two-phase configuration, we examine the principal peak, which in this example is at $k \approx 6 \text{ \AA}^{-1}$.
While the initial configuration contains a \emph{perfect} crystalline half which produces a strong signal, a fully crystalline system will exhibit peaks that are \emph{extensive}, regardless of fluctuations, reflecting global crystalline order.
As such, the principal peak approximately doubles, since the solid now occupies the entire supercell rather than just half.
In the liquid phase, where there is no such global order, all peaks disappear on this scale.

\section{Volume and energy changes upon melting}

\begin{figure}
    \includegraphics[width=\textwidth]{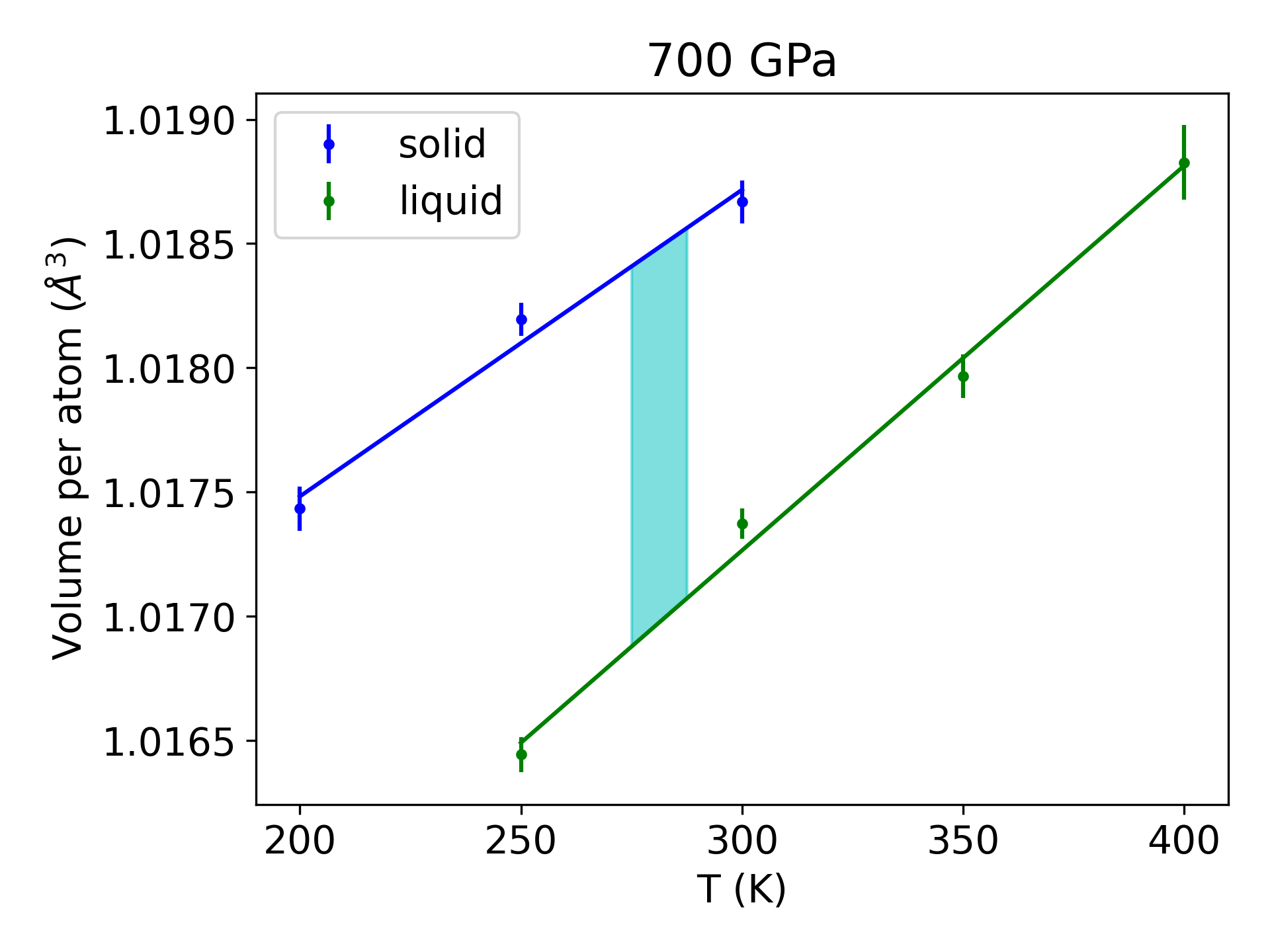}
    \caption{\label{fig:volume_example}
    Volumes of the solid and liquid at various temperatures and 700 GPa.
    A line is fit through these measurements, and from these lines the change in volume can be estimated over the range of temperatures corresponding to the melting point, indicated by the shaded region.
    Note that the liquid is clearly smaller (more dense) than the solid.
    }
\end{figure}

We estimated the change in volume and energy upon melting by performing additional single-phase simulations.
Shown in Figure \ref{fig:volume_example} is an example of how we calculate this change for the volume.
The volumes of the solid and liquid are measured at fixed pressure (700 GPa, in this example) and at various temperatures around and through the melting point.
Over these temperature ranges the change is assumed to be linear, so we fit a line through our data.
This line is then used to interpolate and estimate the change in volume over a range of temperatures, indicated by the shaded region, since we do not have an exact melting point.

Uncertainties are estimated using a jackknife procedure \cite{jackknife}.
A ``best'' estimate $\Delta V$ is produced using all of the available data and averaged over the shaded region.
One of the data points $i$ is then removed, affecting the fit of the lines and resulting in a generally different change $\Delta V_i$.
This is repeated for all data points, and the uncertainty is estimated as
\begin{equation}
    \label{eq:jackknife}
    \sigma_{\Delta V} \approx \sqrt{\sum_i \left( \Delta V - \Delta V_i \right)^2}
\end{equation}
the variation among all of these resampled estimates.

\section{Melting lines}

To approximate our melting lines, we combined the results of our two-phase simulations and Clausius-Clapeyron estimates of $dT/dP$.
Each melting line $T_m (P)$ has the polynomial form
\begin{equation}
    \label{eq:polynomial}
    T_m = c_0 x + c_1 x + c_2 x^2 + c_3 x^3 + c_4 x^4
\end{equation}
where $x \equiv P - 500$ GPa, and $T_m$ is in K.
The coefficients are determined by fitting the derivative while also respecting the bounds from the two-phase results.
\begin{table*}
    \caption{\label{tab:coefficients}
        Coefficients of \eqref{eq:polynomial} specifying our melting lines
    }
    \begin{ruledtabular}
        \begin{tabular}{lcccc}
            & H & D & Classical \\
            \colrule
            $c_0$ & 289.86 & 330.50 & 362.65 \\
            $c_1$ & $-8.4948 \times 10^{-3}$ & $5.7406 \times 10^{-2}$ & 0.22785 \\
            $c_2$ & $-3.3142 \times 10^{-4}$ & $-4.2318 \times 10^{-5}$ & $-4.9028 \times 10^{-4}$ \\
            $c_3$ & $3.9325 \times 10^{-7}$ & $-1.3963 \times 10^{-6}$ & $-5.0295 \times 10^{-7}$ \\
            $c_4$ & 0 & $2.4172 \times 10^{-9}$ & $1.7652 \times 10^{-9}$ \\
        \end{tabular}
    \end{ruledtabular}
\end{table*}
These coefficients for our three lines are given in Table \ref{tab:coefficients}.

\bibliography{bibliography.bib}